\documentstyle[12pt]{article}
\def\beq{\begin{equation}}
\def\eeq{\end{equation}}

\begin{document}
\begin{titlepage}
\begin{center} {\Large \bf Theoretical Physics Institute\\
University of Minnesota}\end{center}
\begin{flushright} TPI-MINN-93/91-T\\
UMN-TH-1212/93\\
August 1993 \end{flushright}
\vspace {0.05in}
\begin{center} {\LARGE Infinite renormalization of $\theta$-term\\
and Jarlskog invariant for $CP$-violation}
\end{center}
\vspace {0.05in}
\begin{center} {\bf I.B.Khriplovich}\footnote{e-mail address:
KHRIPLOVICH@INP.NSK.SU}\\
Budker Institute of Nuclear Physics\\
630090 Novosibirsk, Russia\\
and\\
{\bf A.I.Vainshtein}\footnote{e-mail address:
VAINSHTEIN@MNHEP.HEP.UMN.EDU}\\
Theoretical Physics Institute\\
University of Minnesota, 116 Church St. SE\\
Minneapolis, Minnesota  55455\\
and\\
Budker Institute of Nuclear Physics\\
630090 Novosibirsk, Russia
\vspace {0.1in}
\end{center}
\begin{center} {\bf Abstract}
\end{center}

     The logarithmic ultraviolet divergence in the $\theta$-term induced
by electroweak interactions in the Kobayashi-Maskawa model of
$CP$-violation, found by Ellis and Gaillard, is discussed. We relate it to the
Jarlskog $CP$-odd invariant of quark mass matrix. The divergence arises in
14th order in Higgs coupling as well as in 12th order plus 2nd order in
$U(1)$ gauge boson coupling.
\end{titlepage}

     The purpose of this note is to comment on the connection between
the infinite renormalization of the $\theta$-term found by Ellis and
Gaillard\cite{eg} in the Kobayashi-Maskawa (KM) model of $CP$-violation and
the Jarlskog invariant\cite{ja} of quark mass matrix for $CP$-violating effects
in this model.

It was first noted by Wilczek\cite{wi} that the parameter $\theta$ in the
structure
\beq
\theta\,\frac{g_S^2}{32\pi^2}\,G_{\mu\nu}^a \tilde{G}_{\mu\nu}^a
\label{1}
\eeq
can get infinitely renormalized due to $CP$-violating electroweak
interactions. Then it was shown in ref.\cite{eg} that such infinite
renormalization can occur only starting from 14th order in the
weak coupling $g_W$. It looks only natural that operator (\ref{1}) of
dimension 4 gets infinitely (logarithmically) renormalized. Amazing is
such a high order of perturbation theory where the infinite
renormalization shows up.

On the other hand due to the GIM mechanism the very existence of this
ultraviolet logarithmic divergence may look strange. Indeed in a
renormalizable gauge the ultraviolet divergence of any loop contributing
to the operator (\ref{1}) cannot be stronger than logarithmic one. One
may expect that the GIM mechanism leads to cancellation of logarithmic
divergencies in the $\theta$-term. Such an improvement of the
ultraviolet behaviour takes place indeed. In particular the order
$g_W^4\,g_S^2$ correction to the $\theta$-term turns out to be
finite even in the local limit of the four-fermion interaction\cite{kh}.

The crucial observation of ref.\cite{eg} is that the $\theta$-term
divergence arises in a renormalizable gauge not due to the
$W^{\pm}$-exchange, but due to the Higgs one. The Higgs coupling with
fermions has the form
\beq
\phi^0\,(\bar{D}_L\,H_D\,D_R\,+\,\bar{U}_R\,H_U^{\dag}\,U_L)\,\,+
\,\,\phi^+\,(\bar{U}_L\,H_D\,D_R\,-\,\bar{U}_R\,H_U^{\dag}\,D_L)\,+\,h.c.
\label{2}
\eeq
Here $\phi^0,\,\phi^+$ are the Higgs field components,
\begin{eqnarray}
U=\left( \begin{array}{c}u\\c\\t \end{array} \right),\;\;\;\;\;\;\;\;\;\;
D=\left( \begin{array}{c}d\\s\\b \end{array} \right)
\label{3}
\end{eqnarray}
are the sets of up and down quarks, $H_U,\,H_D$ are the Higgs coupling
matrices directly related to the mass matrices $M_U,\,M_D$ of U- and
D-quarks:
\beq
H_Q\,=\,\frac{g_W}{2m_W}\,M_Q,\;\;\;\;\;\;\,Q=U,D.
\label{4}
\eeq
The coefficient $g_W/2m_W$ in eq.(\ref{4}) has the well-known meaning
of the inverse vacuum expectation value of the neutral Higgs field.

Let us emphasize that GIM cancellation occurs differently
in the Higgs exchange and in the $W$-boson one (in a renormalizable
gauge). In more common $W$ case it is due to the subtraction of
fermionic propagators with different masses which improves indeed
ultraviolet behaviour of loops. For the Higgs exchange the fermion
mass differences discussed can be organized directly as a proper
combination of the Yukawa coupling constants. The difference in the
mechanism can be conveniently formulated in the symmetric (unbroken)
phase. In the first, $W$ case expansion in masses can be interpreted
as an emission of neutral Higgs quanta with vanishing
4-momentum. Meanwhile in the second case we deal with the exchange by
virtual Higgs quanta and the integration over their 4-momenta, i.e.,
with true radiative corrections in the Higgs coupling.

The necessary combination of the Higgs coupling constants is fixed in
fact by the $CP$-odd nature of the $\theta$-term. It coincides (up to a
factor)  with the $CP$-odd invariant of the Higgs coupling matrices
(mass matrices) found in ref.\cite{ja}:
\beq
J\,=\,Im\,det[H_U H_U^{\dag},\,H_D H_D^{\dag}]\,=
\,(\frac{g_W}{2m_W})^{12}Im\,det[M_U M_U^{\dag},\,M_D M_D^{\dag}].
\label{5}
\eeq
It is convenient to use the following representation of the mass
matrices:
\beq
M_Q\,=\,e^{i\eta_Q}\,L_Q M_Q^d R_Q^{\dag}
\label{6}
\eeq
where $M_{U,D}\,=\,diag(m_{u,d}, m_{c,s}, m_{t,b})$ are positive
definite diagonal mass matrices of up and down quarks, $L_Q$ and $R_Q$
are unitary unimodular matrices. Two more parameters in
eq.(\ref{6}) are the $U(1)$ phases $\eta_{U,D}$. The CKM matrix $V$ is
then
\beq
V\,=\,L_U L_D^{\dag}
\label{7}
\eeq
The perturbation expansion of the $\theta$-term in the Higgs coupling
(in renormalizable gauge) dictates just the form (\ref{5}) of the
invariant. Its explicit relation to the $CP$-odd invariant of the CKM
matrix $V$ is\cite{ja}
\beq
J\,=\,(\frac{g_W}{2m_W})^{12}(m_t^2-m_u^2)(m_t^2-m_c^2)(m_c^2-m_u^2)
(m_b^2-m_d^2)(m_b^2-m_s^2)(m_s^2-m_d^2)\,j\;
\label{8}
\eeq
\beq
j\,=\,Im\,V_{cs}V_{us}^*V_{ud}V_{cd}^*
\label{8a}
\eeq
In the standard parametrization\cite{pd} $j=c_{12}c_{23}c_{13}^2 s_{12}
s_{23} s_{13} \sin{\delta_{13}}$.
Let us note that to derive eq.(\ref{8}) it is convenient to use the
simple relation $detC\,=\,(1/3)Tr(C^3)$ for a traceless matrix
$C\,=\,[M_U M_U^{\dag},\,M_DM_D^{\dag}]$ and the observation\cite{ja}
that the imaginary part of the product $V_{ij}V_{kj}^*V_{kl}V_{il}^*$
either coincides (up to a sign) with $j$ in eq.(\ref{8a}) or vanishes.

The necessity of using the combination $M_Q M_Q^{\dag}$ in
eq.(\ref{5}) is determined by the independence on right-handed
rotations $R_Q$. Simultaneously it guarantees the independence on
$U(1)$ phases $\eta_Q$. Certainly one could use arbitrary powers of
$M_Q M_Q^{\dag}$  in the commutator (\ref{5}) (see ref.\cite{ja}).
Being interested in the lowest possible order in the Higgs coupling we
are limiting ourselves just by the first power.

Let us come back now to the original problem of the possible
logarithmic divergence in the induced $\theta$-term, $\theta^{ind}$.
{}From the $CP$-odd nature of the effect it is clear that it should be
proportional to the invariant (\ref{5}),
\beq
\theta^{ind}\,\propto\,J\,\log{M_{uv}}
\label{9}
\eeq
where $M_{uv}$ is an ultraviolet cut-off. Eq.(\ref{9}) implies
immediately that the divergence might appear in 12th order in the
Higgs coupling. Let us note that in the study of the logarithmic divergence
(\ref{9}) it is sufficient to consider the unbroken (symmetric) phase of the
theory, with massless fermions and vector bosons.

Actually the divergence (\ref{9}) is absent even in the 12th order. This
is due to the invariance of the fermion-Higgs interaction (\ref{2})
under the substitution\footnote{The invariance (\ref{10}) is a leftover
of a global $SU(2)_R$ symmetry exact in the limit $H_U=H_D$.}
\begin{eqnarray}
\phi=\left( \begin{array}{c}\phi^+\\ \phi^0 \end{array} \right)\;
\leftrightarrow \;
\phi^c=\left( \begin{array}{c}\bar{\phi^0}\\ -\bar{\phi^-} \end{array} \right),
\;\;\;
U_R\,\leftrightarrow\,D_R,\;\; H_U\leftrightarrow H_D.
\label{10}
\end{eqnarray}
However, the r.h.s. of eq.(\ref{9}) is odd under this transformation,
$J\rightarrow -J$. Therefore to the 12th order the effect still vanishes.

One way to circumvent this cancellation between up and down quarks was
found in ref.\cite{eg}. It is to take into account the interaction with the
$U(1)$ neutral vector boson (linear combination of a photon and $Z$-boson)
coupled to the weak hypercharge $Y$ which is different for right-handed
$U$ and $D$-quarks, $Y(U_R)=4/3, Y(D_R)=-2/3$. Then the logarithmic
divergence in the $\theta$-term arises in 14th order, 12th in the Higgs
coupling and 2nd in the vector boson one:
\beq\label{11}
\theta^{ind}\,\propto\,(g'_W)^2\,[Y^2(U_R)-Y^2(D_R)]\,J\,\log M_{uv}
\eeq
where $g'_W$ is the $U(1)$ gauge coupling constant. The expression
(\ref{11}) is obviously invariant under the transformation (\ref{10}).
We would like to point out here that the effect arises also without going
beyond the fermion-Higgs sector, to 14th order in the Higgs coupling:
\beq\label{12}
\theta^{ind}\,\propto\, Tr(H_U H_U^{\dag}-H_D H_D^{\dag})\,J\,\log M_{uv}.
\eeq
It is due to the extra Higgs exchange instead of a vector boson one.

We wish to make few comments more. Let us consider Feynman diagrams
generating the effective $\theta$-term (see fig.1). Here dashed lines
denote external gluons, solid lines refer to fermions, Higgs and $U(1)$
boson exchanges are confined in rectangular blocks. Strictly speaking, the
$\theta$-term (\ref{1}), being a total derivative, vanishes in the
effective action after integrating over 4-coordinates. To avoid the
complication one can use, e.g., the external field technique.

The induced $\theta$-term is split into two parts due to diagrams 1a and 1b
respectively
\beq
\theta^{ind}=\theta^{ind}_a + \theta^{ind}_b.
\label{12a}
\eeq
In fig.1a both gluons are coupled to the same fermion while in fig.1b
their vertices are separated by electroweak exchanges. In fact, diagram 1a
gives nothing else but the $\theta$-term generated by the induced $CP$-odd
phases of fermion mass matrices,
\beq
\theta^{ind}_a=\eta^{ind}_U + \eta^{ind}_D
\label{13}
\eeq
where $\eta_{U,D}$ are defined by eq.(\ref{6}). This is just that
contribution to $\theta^{ind}$ which was discussed in ref.\cite{eg}.

There are some subtle points in the calculation of diagram 1a. Namely, the
last integration (over the momentum of a fermion coupled to gluons) is of
infrared nature. It should be regularized by the explicit introduction of
the fermion mass $m_F$ so that this integral is dominated  by momenta
$p\sim m_F$.

There are no complications of the kind in $\theta^{ind}_b$, the part of
$\theta^{ind}$ due to the ``irreducible'' diagram 1b which is completely
dominated by high momenta $p\sim M_{uv}$. So, the above analysis leading
to formulae (\ref{11}) and (\ref{12}) is directly applicable to this
contribution to $\theta^{ind}$.

Coming back to the $CP$-odd part of the fermion mass operator (rectangular
block in diagram 1a), it is interesting to note that this part arises
already to 8th order in the Higgs coupling. For instance, $CP$-odd mass
operator of $u$-quark can be shown to have the following structure:
\beq
\eta_u=\frac{m_u^{CP-odd}}{m_u}\,\propto\,(\frac{g_W}{2m_W})^8
(m^2_t-m^2_c)(m^2_b-m^2_s)
(m^2_s-m^2_d)(m^2_d-m^2_b)j \log M_{uv}
\label{14}
\eeq
where $j$ is the $CP$-odd invariant (\ref{8a}) of the $CKM$ matrix. The
derivation of eq.(\ref{14}) is similar to that of eq.(\ref{8}). In
terms of the initial invariant $J$ (see eq.(\ref{5})) expression
(\ref{14}) is singular in the mass differences:
\beq
\eta_u \,\propto\,(\frac{2m_W}{g_W})^4 \frac{J \log
M_{uv}}{(m^2_u-m^2_t)(m^2_c-m^2_u)}
\label{15}
\eeq
It looks contradicting to our above analysis where we assumed a
proportionality to $J$ without extra singularities in the Higgs coupling.
The explanation of the singularity in eq.(\ref{15}) is as follows. It
arises in the limit of degenerate $U$-quarks. But when we are dealing with
degenerate states, it is only the sum over them which should be
nonsingular in the parameter of degeneracy\cite{ln}. Indeed, adding up the
8th order contributions of $U$-quarks into $\theta^{ind}$ (see eqs.(\ref{13})
and (\ref{6})) one gets zero:
\beq
\eta_U^{ind}=\eta_u^{ind}+\eta_c^{ind}+\eta_t^{ind}=0.
\label{16}
\eeq
Of course, the similar effect takes place for $D$-quarks. Moreover,
the same cancellation persists in 10th order in the Higgs coupling where
the singularity, say, of $\eta_u^{ind}$ is one power less than in
eq.(\ref{15}).
No singularity at all occurs in 12th order, $\eta_U\,\propto\,J\log M_{uv}$.
However, in this order the cancellation takes place between $U$- and
$D$-quarks, $\eta_U^{ind}+\eta_D^{ind}=0$.
In this way the part of $\theta$ generated by the mass matrix,
$\theta_a^{ind}$, has the same form (\ref{11}) or (\ref{12}) as the
``irreducible'' one, $\theta_b^{ind}$.

In conclusion let us note that at any reasonable value of the
ultraviolet cut-off parameter $M_{uv}$ the discussed 14th order
contribution to the $\theta$-term is absolutely negligible numerically.
The leading finite correction to the $\theta$-term arises in the
$g_W^4 g_S^2$ order\cite{kh}. It is ``irreducible'', i.e. it is not
generated by the CP-odd part of fermion mass operator.

\vspace{0.5cm}

\begin{flushleft}
{\bf Acknowledgments} \end{flushleft}

We are grateful to V.A. Miransky, M.E. Pospelov and P. Ramond
for stimulating discussions. The work was supported in
part by Department of Energy under the grant DOE-AC02-83ER40105.

\vspace{1cm}
{\bf Figure captions}\\
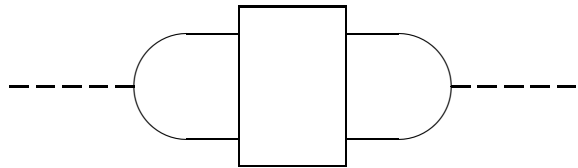
Fig. 1. Diagrams generating $\theta$-term
\newpage
\begin{figure}
 \begin{picture}(460,180)
  \put(200,110){\oval(80,40)[l]}
  \put(240,110){\oval(80,40)[r]}
  \put(200,130){\line(1,0){40}}
  \put(200,80){\framebox(40,20)}
  \multiput(113,110)(10,0){5}{\line(1,0){7}}
  \multiput(280,110)(10,0){5}{\line(1,0){7}}
  \put(215,50){(a)}
             \end{picture}

 \begin{picture}(460,180)
  \put(200,140){\oval(80,40)[l]}
  \put(240,140){\oval(80,40)[r]}
  \put(200,110){\framebox(40,60)}
  \multiput(113,140)(10,0){5}{\line(1,0){7}}
  \multiput(280,140)(10,0){5}{\line(1,0){7}}
             \put(214,80){(b)}
  \put(213,0){\makebox(0,0){Figure 1}}
             \end{picture}

\end{figure}

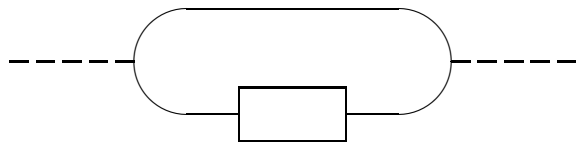
\end{document}